\documentclass[a4paper,11pt]{article}
\pdfoutput=1
\usepackage{graphicx}
\usepackage{bm}
\usepackage{epsf}
\usepackage{rotating}
\usepackage{epsfig,graphics,rotate,color}
\usepackage{color}
\usepackage{array}
\usepackage[utf8]{inputenc}
\usepackage[T1]{fontenc}
\usepackage[caption=false]{subfig}
\usepackage[bottom]{footmisc}
\usepackage{adjustbox}
\usepackage{multirow}

\usepackage[utf8]{inputenc}
\usepackage{amsmath}
\usepackage{amsfonts}
\usepackage{amssymb}
\usepackage{lineno}

\pdfoutput=1 
\usepackage{jinstpub} 

\newcommand{\beq}{\begin{eqnarray}}
\newcommand{\eeq}{\end{eqnarray}}

\def\Cdnb{12} 
\def\Cdxs{2.1\times 10^4} 
\def\CdQv{9.04} 

\def\AbsWat{470}

\def\EmiWat{460}

\def\QDSY{4000}

\def\PLQYoneone{15.6}\def\PLQYoneoneE{1.2}
\def\PLQYonethr{ 8.8}\def\PLQYonethrE{0.8}
\def\PLQYtwotwo{ 8.8}\def\PLQYtwotwoE{1.5}
\def\PLQYtwothr{14.5}\def\PLQYtwothrE{0.7}
\def\PLQYthrthr{ 9.5}\def\PLQYthrthrE{0.5}

\def\Dhoneone{52.2}\def\DhoneoneE{0.3}
\def\Dhonethr{48.7}\def\DhonethrE{0.1}
\def\Dhtwotwo{60 }\def\DhtwotwoE{1 }
\def\Dhtwothr{55.7}\def\DhtwothrE{0.3}
\def\Dhthrthr{71.6}\def\DhthrthrE{0.3}

\def\Zetaonethr{-13.8}\def\ZetaonethrE{0.8}
\def\Zetatwotwo{-24 }\def\ZetatwotwoE{2 }
\def\Zetatwothr{-47 }\def\ZetatwothrE{2 }
\def\Zetathrthr{-24.6}\def\ZetathrthrE{0.7}

\def\SizeDLSTC{25}
\def\SizeDLSTE{9}
\def\SizeDLSTS{11}
\def\SizeDLSWC{71.6}
\def\SizeDLSWE{0.3}
\def\SizeDLSWS{26}
\def\SizeTEMTC{6.4}
\def\SizeTEMTE{1.2}
\def\SizeTEMPT{6.4}
\def\SizeTEMPTE{1.0}
\def\SizeEmpiC{5.22}
\def\SizeEmpiE{0.01}

\def\daqwindow{4}

\title{Water-based Quantum Dots Liquid Scintillator for Particle Physics}
\author{
M.~Zhao, 
M.~Taani, 
J.~Cole, 
B.~Crudele\footnote{Now at University College London, London, UK.}, 
B.~Zou, 
N.~Bhuiyan, 
E.~Chowdhury, 
Y.~Duan, 
S.~Fekri\footnote{Now at University of Oxford, Oxford, UK.}, 
D.~Harvey\footnote{Now at University of Queensland, Brisbane, Australia.}, 
D.~Mitra\footnote{Now at Virginia Tech, Blacksburg, VA, USA.}, 
O.~Raz\footnote{Now at University of Chicago, Chicago, IL, USA.}, 
A.~Thompson\footnote{Now at Royal Holloway, University of London, Egham, UK.}, 
T.~Katori, 
and A.~Rakovich 
}
\affiliation{Department of Physics, King's College London, WC2R 2LS London, UK.}

\emailAdd{teppei.katori@kcl.ac.uk, aliaksandra.rakovic@kcl.ac.uk}

\abstract{Liquid scintillators are typically composed from organic compounds dissolved in organic solvents. However, usage of such material is often restricted due to fire safety and environmental reasons. Because of this, R\&D of water-based liquid scintillators is of extreme relevance; yet, no such scintillators have been made commercially available as yet. Here, we investigate an alternative, water-based quantum dots liquid scintillator. Pre-determined and controllable optical properties of the quantum dots, as well as the existence of large libraries of established protocols for their dispersion in aqueous solutions, make them an attractive option for nuclear and particle physics applications.
We characterize the optical properties of water-based quantum dots liquid scintillator and find that most of its optical properties are preserved upon quantum dots' phase transfer into water, through the addition of an oleic acid hydrophilic layer. Using the developed scintillator, the time and charge responses from atmospheric muons are measured, highlighting the practical viability of water-based quantum dots liquid scintillators for nuclear and particle physics, special interest on neutrino physics.
}

\keywords{reactor neutrino, water-based liquid scintillator, quantum dots, cadmium sulfide, oleic acid coating}

\begin{document}
\maketitle
\flushbottom

\section{Introduction}
\label{sec:Introduction}

Since their first experimental observation in glass matrices~\cite{Ekimov1982soviet}, nano-sized crystals of semiconducting materials, \textit{aka} Quantum Dots (QDs), have been the focus of research in many different areas. Their superior and size-dependent optoelectronic properties make them attractive for many applications, including nanomedicine~\cite{Rakovich2018qdmed}, bioimaging~\cite{Bilan2016qdbio}, photocatalysis~\cite{Sun2023qdcat}, photovoltaics~\cite{Lu2020QDsolar}, and lighting and display applications~\cite{Tian2023qdlight}. The field of QD-based photonics has advanced rapidly since their first realization, with some examples of translation into commercial products occurring within the last decade~\cite{commercial-link} which highlight the suitability of QDs for technological applications.

The most attractive properties of QDs for photonics applications are their high extinction cross-sections and photoluminescence quantum yields (PLQY), both of which are typically higher or comparable to molecular equivalents~\cite{QDphoton}. In addition, they have larger physical cross-sections, which increases the probability of their interaction with charged particles and subsequent generation of photon signals through direct scintillation. Direct scintillation has been observed from QD-doped solid materials under irradiation~\cite{QDradiation,WANG2010186}. Therefore, we expect that direct scintillation through energy exchange with other radiations, for example minimally ionizing particles (MIPs) such as cosmic rays (atmospheric muons), are equally possible.
  
The last few decades of research into the properties of colloidal QDs established
a large library of surface modification protocols~\cite{QDsurface}, including molecular scintillators or dopant-containing molecules, which provide a route to introduce additional components into the system without adverse effects to solvent transparency.
Using this approach, QDs can simultaneously act as direct liquid scintillators and as wavelength-shifters for the molecular scintillators or dopant molecules attached to their surfaces, where the energy is transferred resonantly between the two.
This presents a new prospect of being able to improve light yields
of the solution without modifying the overall concentration of solute in the solution.
QDs of appropriate surface functionalisations are also directly dispersible in water~\cite{C2CS35285K}, allowing them to act as a water-based quantum dots (WbQD) liquid scintillator.

Here we are investigating the possibility of utlizing QDs for particle physics applications~\cite{Winslow:2012ey,Blanco:2022cel,Doser:2022knm}. In particular, we are interested in the possibility of WbQDs liquid scintillator in nuclear fission reactor neutrino detection. Water is a safe material to use near the fission reactor core, and such detector can be a new type of near-field fission reactor monitoring detector~\cite{Yeh:2011zz,Alonso:2014fwf,Theia:2019non,Caravaca:2020lfs,Kneale:2022vpw}. Typical compositions of the QDs include elements such as cadmium, selenium, and tellurium; and whilst some of them pose environmental concerns~\cite{CdToxic1, CdToxic2}, all of them have high neutron capture cross-sections, which opens a possibility to tag neutrons. Hence, WbQDs can act as conventional liquid scintillators with neutron detection ability without requirements for doping with neutron capturing materials, in contrast to the proposed gadolinium-doped water-based liquid scintillator~\cite{Kneale:2022vpw}.

In preparing the WbQD liquid scintillator, the parameters of interest were (1) the photoluminescence quantum yield (PLQY), (2) the transparency, (3) the emission spectra, (4) concentration and (5) the stability of the scintillating liquid. The PLQY, defined as the ratio of photons emitted per photons absorbed by the fluorophore, has to be high enough to be able to efficiently resolve the incoming particle energy. However, high-PLQY colloidal QDs are typically produced in organic solvents and are therefore not stable in aqueous solutions as-synthesized. It is possible to encapsulate such QDs in an amphiphilic layer \textit{via} a process called phase transfer, which facilitates their dispersion in water~\cite{oleic}; however, such processes typically result in a diminished PLQY of the QDs and increased scattering by the colloids. Nonetheless, the resulting dispersions are shown to be sufficiently transparent for the emitted photons to reach the photodetectors without being re-absorbed.

Regarding the emission properties of QDs, the vast majority of photodetectors used for radiation detection have maximum quantum efficiencies (QE) in the blue end of the spectrum; therefore, the selection of QDs' composition and size needs to be such as to yield emission in this spectral range. QDs with Cd-based compositions and small core sizes seem appropriate for proof-of-principle demonstrations. Furthermore, and although not explored in this work, cadmium has a high-neutron capture cross-section and Cd-based cores of our QDs open a possibility to employ them as neutron capture materials, allowing tagging of neutrino interactions through the inverse beta-decay processes~\cite{Bernstein:2019hix}. Regardless of the application, the concentration of QDs in the scintillating liquid needs to be high enough to facilitate sufficient muon interactions or neutron capture events for experimentation, whilst low enough to maintain the optical transparency of the liquid. As such, there will be an optimum range for QD concentration in WbQD scintillating liquids.

Finally, for application purposes, QDs must remain colloidally and functionally stable for relatively long periods of time. Whilst it is known that phase-transferred QDs might gradually loose their stability over time and that routes to mitigate this do exist~\cite{QD-phase-transfer1, QD-phase-transfer2, QD-phase-transfer3, QD-phase-transfer4}, little is known about the effect of long-term exposure to radiation on the QDs' stability and functionality. Although radiation responses of QDs were studied in the past~\cite{QDradiation,WANG2010186}, these were for low energy particles and not MIPs such as cosmic rays. As such, measuring MIP responses and therefore demonstrating the scintillating property of WbQDs liquid scintillator - the subject the work presented here - is the first step towards the utilization of WbQDs as particle scintillators.

In Section~\ref{sec:sample}, we describe the selection and preparation of WbQD scintillating liquids used in this work. Section~\ref{sec:optics} discusses the optical and physical properties of the prepared liquids, as well as their long-term stability. In Section~\ref{sec:cosmic}, we study the cosmic ray responses from WbQD samples to evaluate their potential as a liquid scintillator for a radiation detector. The results are discussed and critically evaluated in Section~\ref{sec:discussions}. Finally, Section~\ref{sec:conclusion} summarizes key findings and proposes directions for future studies.

\section{Quantum dots sample preparation
\label{sec:sample}}

\subsection{Quantum dots selected}

To meet the required characteristics discussed in the previous section, cadmium sulfide (CdS) core zinc sulfide (ZnS) shell type coating quantum dots were selected for the experiments. There were numerous reasons behind this selection.
Firstly, the CdS composition of the QD core facilitates shorter wavelengths of QD fluorescence, compared to \textit{e.g.} the more popular CdSe or CdTe QDs.
Secondly, blue-emitting CdS QDs of relatively high PLQY ($\sim$50\%) were available for purchase commercially ~\cite{Aldrich}, although these QDs could not be dispersed in water directly, requiring further surface modification to transfer them to aqueous environments (as detailed in sec.~\ref{sec:Methods}).
Thirdly, the manufacturer stated the shelf-life of these QDs as being three years, which was sufficient for our purposes due to the short time span of the experiment.
And finally, the cadmium isotope $^{113}$Cd has a natural abundance of $\Cdnb\%$, and a relatively high neutron capture cross-section of $\Cdxs$~barn~\cite{Liu:2016mjp}, with a Q-value $\CdQv$~MeV.

The CdS/ZnS QDs satisfied many of the criteria posed in the previous section, and it was expected that these QDs may allow for particle identification, \textit{i.e.} to distinguish neutrons from other signals with a suitable concentration. This would be especially useful as a low-energy neutrino detector, such as a fission reactor neutrino detector. This property will be studied in future. Here, our target concentration of cadmium is 0.01\% by mass, to match the recent gadolinium doping by the Super-Kamiokande experiment~\cite{Super-Kamiokande:2023xup}. In our case, this corresponds to roughly 10~mg of CdS/ZnS QDs in 80~mL water.

\subsection{Surfactant encapsulation method}
\label{sec:Methods}
Oleic acid (OA) functionalized CdS/ZnS core-shell QDs (emission wavelength $440-460$~nm, $\sim$50\% quantum yield), supplied in solid form, were purchased from Sigma Aldrich~\cite{Aldrich}. Since the as-supplied QDs cannot be directly dispersed in water, QDs were first dispersed in toluene and then phase transferred into water using the surfactant encapsulation method. The procedure for the phase transfer followed closely the previously reported methodolody\cite{oleic}, involving encapsulation of QDs in a double layer of OA, which is known to produce extremely stable QDs. Briefly, 10~mg of CdS/ZnS QDs powder was dispersed in 1.6 mL toluene and the stock solution was left to completely evaporate under vacuum and at room temperature, in the dark. The QDs solids were then dispersed in 16 mL of hexane, to which 320 $\mathrm{\mu}$L of OA was added. The dispersion was then sonicated for 1 min using the Ultrasonic bath XUBA3~\cite{XUBA3}, after which 200 mL of Milli-Q water (Millipore, 16 M$\Omega$-cm) was added to the mixture, whilst being continuously stirred. Subsequent sonication of the mixture over a period of one hour, with occasional manual shaking of the vial (every 3 minutes), resulted in a formation of an emulsion. This was then kept in undisturbed in darkness and at room temperature for 75h, during which time phase separation took place. At this point, QDs should have phase transferred to the water phase (bottom half of the phase separated liquid) and they were removed carefully using a syringe. The dispersion was then centrifuged at 6500 rpm for 15 min, using Eppendorf Centrifuge 5804~\cite{Eppendorf} to cause precipitation of any large QD aggregates. The remaining supernatant was collected and filtered through a 220~nm pore syringe filter, to remove any remaining smaller QD aggregates. We note that, during the above procedure, a small aliquot of toluene-dispersed QDs was set aside for control measurements.

\begin{figure}[ht]
    \centering
\includegraphics[width=1\textwidth]{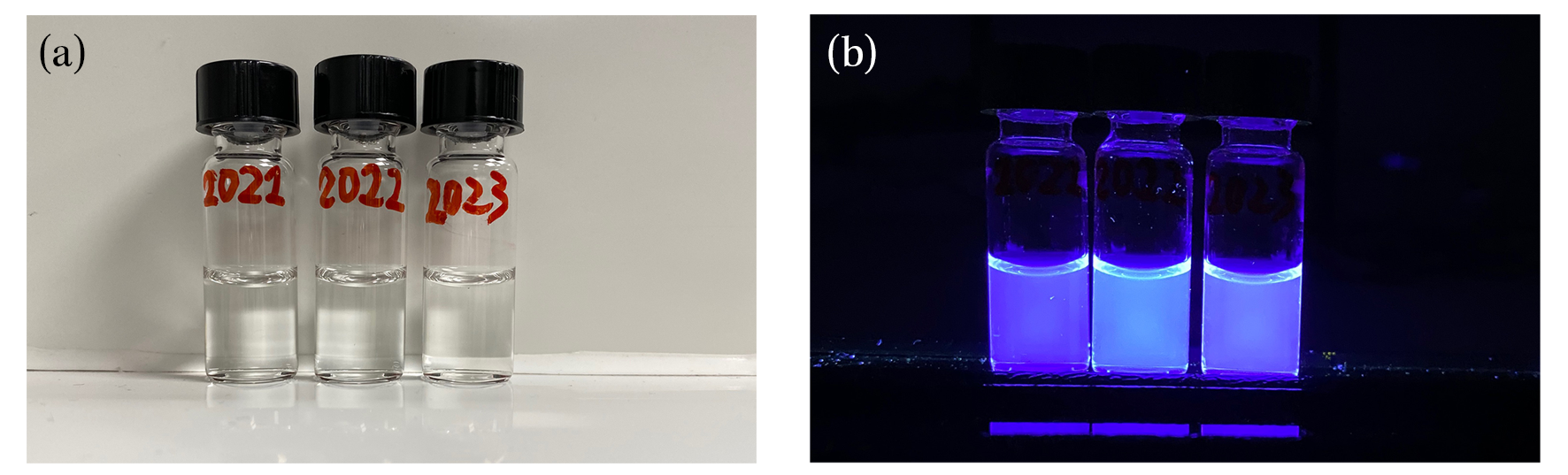}
    \caption{The water-based CdS quantum dots samples, imaged under (a) ambient light and (b) UV lamp. In both images, the sample order is oldest (2021) to newest (2023). The blue color of the solutions visible under UV illumination is due to the fluorescence of CdS quantum dots.}
\label{fig:sample}
\end{figure}

Three samples were prepared in this manner, one in each of the years of 2021, 2022 and 2023, and labelled accordingly. Figure~\ref{fig:sample} shows the three samples under natural and UV illumination (recorded in August of 2023). The blue color of solutions under UV illumination (in panel (b)) is indicative of QD presence in the dispersions and therefore of the successful completion of the phase transfer procedure on each occasion. It is clear that in all three cases the concentration of QDs is relatively high; yet, and as is evident from panel (a) of Figure~\ref{fig:sample}, all three samples have a high degree of transparency, as per requirements for successful neutrino experiments. In the following result section, we present a representative set of data from the 2023 sample; the properties of 2021 and 2022 samples will be discussed in detail in the section~\ref{sec:aging}, where sample aging is considered. We note that, post-fabrication, all samples were stored in a non-transparent case at $4^\circ$C. Prior to testing, a small aliquot of the sample was equilibrated at room temperature. This was done to ensure that factors such as varied storage conditions and various degrees of sample agitation did not contribute to any observable experimental differences over time.

\section{Physicochemical properties of CdS/ZnS QDs
\label{sec:optics}}
\label{sec:Measurements}

In this section we report the results of the photophysical characterization of the 2023 QD dispersion, including measurements of the QDs' size, surface charge, and their optical properties.

\subsection{Optical properties of CdS/ZnS QDs\label{sec:optical}}
The extinction and photoluminescence (PL) spectra of QD dispersions were recorded using 2600i UV-vis spectrometer from Shimadzu~\cite{Shimadzu} and Cary Eclipse Fluorescence Spectrophotometer~\cite{CaryEclipse} respectively. In our interpretation of the extinction data, we attributed any strong changes in the extinction of the sample to the absorption by the QDs, since they have relatively small scattering cross-sections \cite{xu2020quantification}. However, below the band gap of the QDs, scattering contributions can and do become dominant.

\begin{figure}[ht]
\centering
\includegraphics[width=\textwidth]{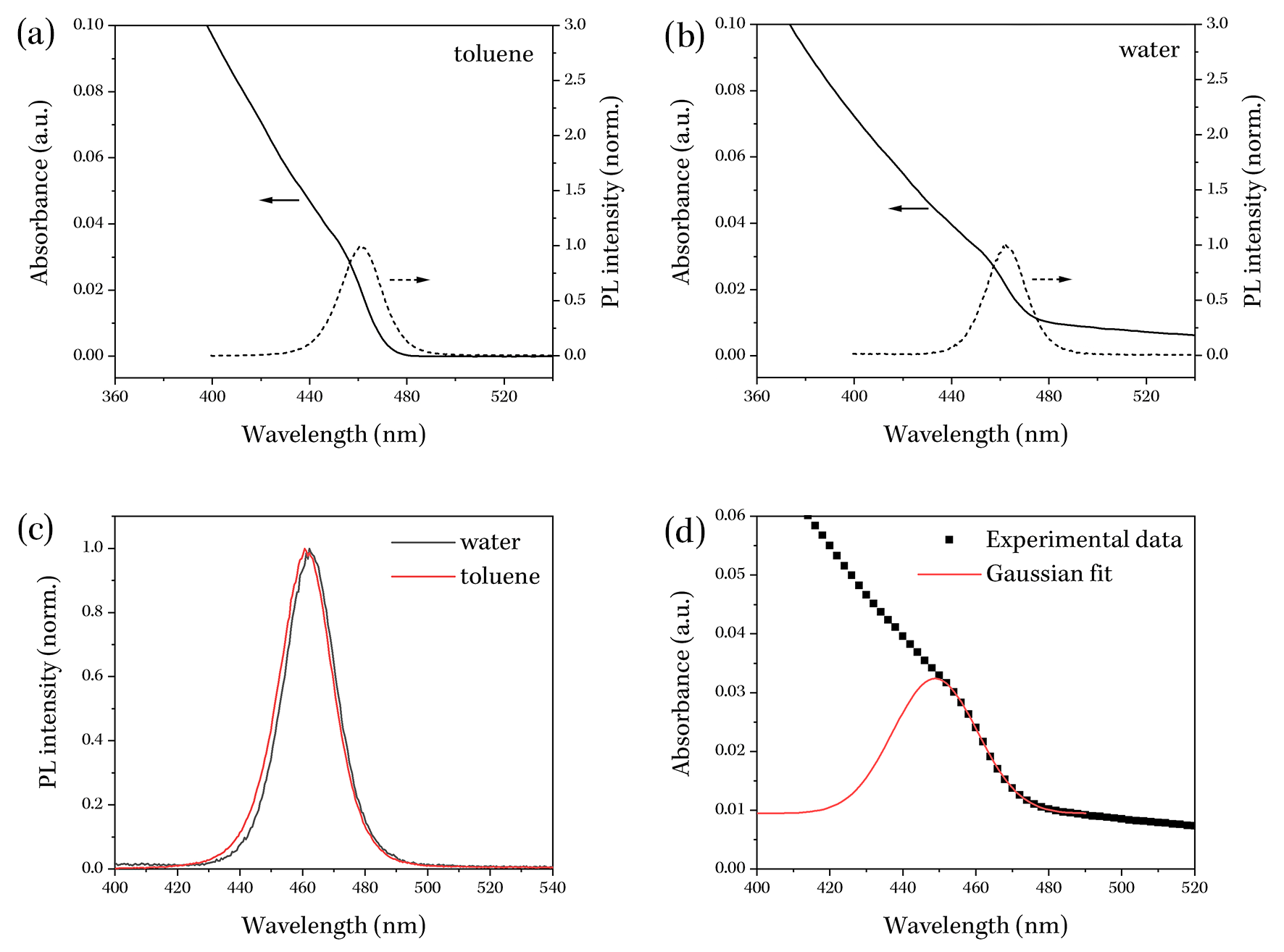}
\caption{\textbf{Optical properties of CdS/ZnS QDs}. Panels \textbf{(a)} and \textbf{(b)} show absorbance (solid lines, left scales in absorbance unit or a.u.) and photoluminescence (dashed lines, right scales in arbitrary normalization) of (a) toluene- and (b) water-dispersed QDs. Panel \textbf{(c)} compares the normalized photoluminescence spectra of the toluene- and water-dispersed QDs samples, both measured at 380 nm excitation. Panel \textbf{(d)} shows the experimental data (black scatter points) for the water-dispersed QD sample and the Gaussian fit to the data, that was used to determine the position and width of the first excitonic peak of the QDs. The $R^2$ value for the fit was 0.99983 and the corresponding $\chi^2$ value was $1.2\times10^{-8}$.}
\label{fig:optical-properties}
\end{figure}

As expected, and as shown in Figs.~\ref{fig:optical-properties}(a) and (b), the absorbance of QDs dispersions in toluene and in water remains negligible at low energies (high wavelengths), but increases rapidly as the energy of the excitation becomes greater than the bandgap of the QDs (at $\sim$470~nm for both). No significant changes to the absorption spectra could be observed for the WbQD liquid as compared to the toluene-dispersed QDs, indicating that no significant chemical or physical changes to the QD core ocurred as a result of the phase transfer process. However, comparison of the extinction spectra of the two samples at wavelengths above the QDs' band edge, it is clear that the phase transfer of QDs to an aqueous environment resulted in an increased Rayleigh's scattering by the QDs - this appears as an elevated but gradually decreasing background at wavelengths larger than 470~nm. This is a typical observation, owing to the size dependency of the Rayleigh scattering process and simply indicates an increased overall size of the QDs upon addition of the surfactant coating that is needed to achieve water-dispersibility~\cite{Rayleigh-article, Rayleigh-book}.

The PL spectra of both WbQD and toluene-based dispersions appeared as Gaussian peaks centred around 460 nm (Fig.~\ref{fig:optical-properties}(a) and (b), excitation at 380 nm). As discussed previously, the observed emission in the blue end of the spectrum is favourable for radiation detection experimentation. The Gaussian shape of the PL spectra is reflective of the size-dispersion of the QDs in the sample, and is very typical and characteristic of colloidal QDs~\cite{QDs-PL-gaussian-article,QDs-PL-gaussian-book}. Of note is also the lack of any significant changes to the emission of the QDs upon phaser transfer, as is evidenced by the normalized PL spectra presented in Fig.~\ref{fig:optical-properties}(c) where only a slight shift in the position of the peak could be observed ($\Delta \lambda < 1~\mathrm{nm}$). This small shift can be attributed to the slight change in the environmental refractive index of the QDs resulting from the addition of an OA shell during the phase transfer process~\cite{QD-PL-shift-in-toluene}.

\begin{figure}[ht]
    \centering
\includegraphics[width=\textwidth]{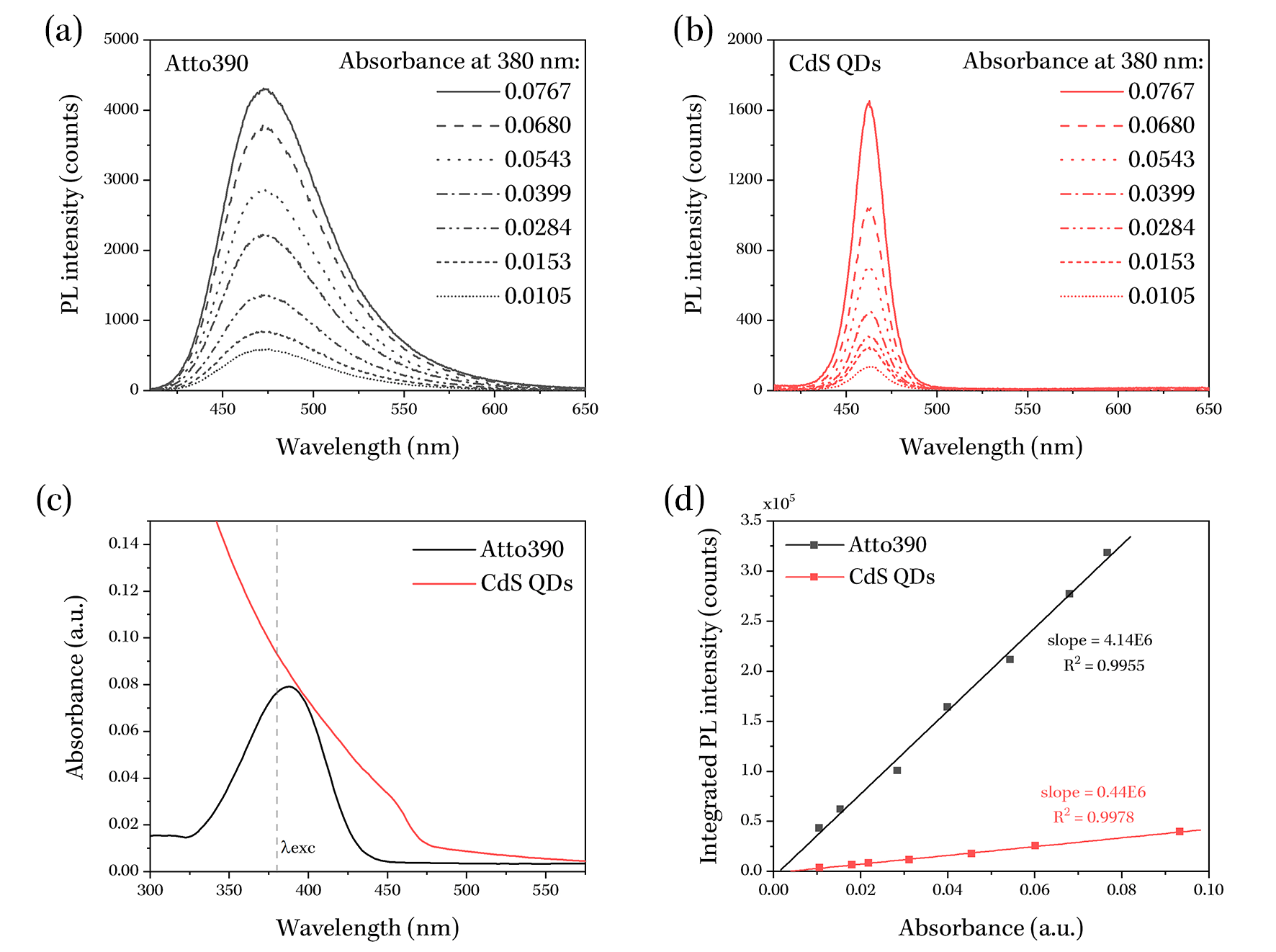}
\caption{\textbf{Determination of PLQY of WbQD liquid.} Panels \textbf{(a)} and \textbf{(b)} show photoluminescence spectra of (a) Atto390 and (b) WbQD samples of seven different concentrations. Absorbance of each sample at the excitation wavelength ($\lambda_{exc}=$380~nm) is stated in the legend of each panel. \textbf{(c)} Extinction spectra of Atto390 (black line) and WbQD samples (red line). The excitation wavelength used for acquisition of photoluminescence spectra, shown in panels (a) and (b), is indicted by the gray dashed line. \textbf{(d)} Integrated photoluminescence intensity of Atto390 (black data) and CdS QDs (red data) as a function of sample absorbance at 380 nm. The experimental data are shown as scatter points. The lines are linear fits to the data; the $R^2$ value of each fit and the slope obtained from it are indicated next to the lines in respective colours. The ratio of the slopes is proportional to the ratio of the quantum yields of the two samples.}
\label{fig:PLQY}
\end{figure}

We further determine the PLQY of the WbQD sample, by comparing its emission properties to those of a reference dye with a known PLQY and emission in a wavelength range that coincides with that of the WbQD sample. Atto390 dye was used for the purpose here, which has a PLQY of $\Phi_{dye}=90\%$ in aqueous environments. Figure \ref{fig:PLQY} summarizes the measurements performed. PL spectra of both the reference dye and the WbQD, diluted to seven different concentrations, were recorded (Figs.~\ref{fig:PLQY}(a) and (b) respectively) using 380~nm as the excitation wavelength in all cases. These concentrations corresponded to a range of sample absorbances, which were recorded using extinction measurements (a representative set is shown in Figure~\ref{fig:PLQY}(c)). The PL spectra were then integrated over the entire spectral range and these integrated values were plotted against corresponding absorbance values at excitation wavelength, as shown in Figure~\ref{fig:PLQY}(d). The resulting sets of data, one for the reference dye and one for the WbQD sample, were then fitted to linear functions of the form
\begin{equation}
\label{eq:linear-fit}
 I(A)=I_0 + K\cdot A
\end{equation}
where \(I(A)\) refers to the integrated PL intensity of the sample at a specific value of sample absorbance, \(I_0\) is an offset/background term. The results of the fitting procedure are summarized in Table~\ref{tab:PLQY-fit} below.

The ratio of the slopes for the WbQD and reference dye solutions are proportional to the ratio of their PLQYs~\cite{PLQY-relative-method}, \textit{i.e.}

\begin{equation}
\label{eq:qy-calculation}
 \Phi_{QD} = \Phi_{dye} \cfrac{K_{QD}}{K_{dye}}.
\end{equation}

Using this relation, and the known value of the PLQY of the reference dye ($\Phi_{dye}=90\%$), the PLQY of our 2023 WbQD liquid was calculated to be
\begin{equation}
\Phi_{QD}=\PLQYthrthr\pm\PLQYthrthrE\%~.
\end{equation}
Clearly, there was an approximately 5-fold reduction of the PLQY of the QY upon their phase transfer from organic to aqueous environment, which is a typical observation for such processes~\cite{QD-phase-transfer-QY1,QD-phase-transfer-QY2}, often attributed to the introduction of surface defects that act as long-lived trap states for the excited states of the QDs~\cite{QD-trap-states}. Nonetheless, the PLQY of the WbQD solution remained sufficiently high for the proposed proof-of-principle demonstrations.

\begin{table}[ht]
\centering
\small
\caption{Results of linear fitting to integrated intensity versus sample absorbance data for the reference dye (Atto390) and 2023 WbQD liquid, shown in Fig.~\ref{fig:PLQY}(d). The data was fitted to Equation~\ref{eq:linear-fit}.}
\begin{tabular}{>{\raggedright}p{0.15\textwidth}>{\centering}p{0.28\textwidth}>{\centering}p{0.3\textwidth}>{\centering\arraybackslash}p{0.14\textwidth}}
  \hline
Sample & Slope, $K$ & Intercept, $I_0$ & $R^2$ value \\
 \hline
Atto390 dye & $(4144\pm124)\times10^3$ & $(-5.25\pm5.98)\times10^3$ & 0.9955 \\ 
2023 WbQDs & $(437\pm9)\times10^3$ & $(-1.30\pm0.44)\times10^3$ & 0.9978 \\
\hline
\end{tabular}
\label{tab:PLQY-fit}
\end{table} 

In addition to information provided above, the absorption spectrum of the QD dispersions was also utilized to estimate the concentration of the prepared dispersions, using a set of empirical equations developed by W. W. Yu \textit{et al.} for aqueous dispersions of CdS QDs~\cite{extinction}. It is important to note that our QDs, composed from a CdS core with ZnS shell, differ from those for which the empirical relations were derived (CdS core only). The addition of the ZnS shell to the CdS core has the effect of passivation of surface defects, leading to improved exciton confinement, quantum yields and photochemical stability of QDs~\cite{QDpassivation1, QDpassivation2}. However, the position of the absorption and emission peaks of the QD solution remain largely unaffected by this change. Since these specific parameters weigh strongly in the method, and only an estimate for the concentration of the sample is required for discussions of results, the use of the method developed by W. W. Yu is justified on this occasion.

The implementation of the method involved the fitting of the measured absorption spectrum of the WbQD solution to a Gaussian function, at the on-set of QD absorption, and as shown in Figure~\ref{fig:optical-properties}(d). Specifically, a subset of the data in the $450-490$ nm spectral range was fitted to a function of the form:

\begin{equation}
\label{eq:gauss-amp}
A(\lambda) = A_0 + A_{max}\cdot \exp{\bigg[-0.5 \bigg( \frac{\lambda-\lambda_c}{w} \bigg)^2 \bigg] },
\end{equation}

\noindent where $A_0$ and $A_{max}$ are the amplitudes of the absorbance background and the Gaussian peak respectively, and $\lambda_c$ and $w$ are the central position and the width of the peak, respectively. As part of the fitting analysis, the software also calculated the full-width at half maximum (FWHM $=27.3\pm0.4$) of the Gaussian peak. The results of the fitting procedure, summarized in Table~\ref{tab:ABS-fit}, were then inputted into the empirical equation of reference~\cite{extinction} to calculate the extinction coefficient, $\varepsilon$ of the WbQD dispersion at the position of the first extinction peak, $\lambda_c$, using the following set of equations:
\begin{equation}
\label{eq:QDsize}
D = (-6.6521\times 10^{-8} ) \lambda_c^{3}+(1.9557\times 10^{-4} ) \lambda_c^{2}-(9.2352\times 10^{-2} ) \lambda_c+(13.29)
\end{equation}
\begin{equation}
\label{eq:extinction}
\varepsilon = 21536 \cdot (D)^{2.3}
\end{equation}
where $D$ refers to the mean diameter of the QDs in the sample.

Furthermore, and as per protocol described in reference~\cite{extinction}, the measured absorbance of the sample at $\lambda_c$ was calibrated to account for the dispersion of QD sizes within the sample according to
\begin{equation}
 \label{eq:abs-adjusted}
 A' = \cfrac{ A\cdot FWHM}{2\cdot FWHM_c},
\end{equation}
where $A$ is the measured absorbance of the sample at $\lambda_c$, $A'$ is the calibrated absorbance and $FWHM_c$ is the half width at half maximum of the first excitonic absorption peak of a typical CdS QD sample ($FWHM_c = 11~\mathrm{nm}$).

\begin{table}[ht]
\centering
\small
\caption{Results of fitting of band-edge absorbance to a Gaussian function shown in Fig.~\ref{fig:optical-properties}(d). The data was fitted to Equation~\ref{eq:gauss-amp} with an $R^2$ value of 0.99983 and a $\chi^2$ value of $1.2\times10^{-8}$.}
\begin{tabular}{>{\raggedright}p{0.08\textwidth}>{\centering}p{0.15\textwidth}>{\centering}p{0.15\textwidth}>{\centering}p{0.15\textwidth}>{\centering}p{0.15\textwidth}>{\centering\arraybackslash}p{0.15\textwidth}}
  \hline
      & $A_0$ & $A_{max}$ & $\lambda_c$ & $w$ & $FWHM$ \\
 \hline
Value & 0.00947 & 0.0229 & 448.8 & 11.6 & 27.3\\ 
Error & 0.00006 & 0.0002 & 0.3 & 0.2 & 0.4\\
\hline
\end{tabular}
\label{tab:ABS-fit}
\end{table} 

Following the described above procedure, the our WbQD sample was determined to a calibrated absorbance of $A'=0.0402 \pm 0.0006$, a mean QD diameter of
\begin{equation}
D=\SizeEmpiC\pm\SizeEmpiE~\mathrm{nm}~,
\label{eq:SizeEmpri}
\end{equation}
and therefore an an extinction coefficient of
\begin{equation}
\varepsilon_{QD}=9.6\times10^5~\mathrm{cm^{-1} M^{-1}}
\end{equation}
at the first extinction peak. Using these two pieces of information, as well as the known path length used in the absorbance measurements ($L=1~\mathrm{cm}$), the concentration of the sample ($C$) could then be determined from the Beer-Lambert law,
\begin{equation}
 \label{eq:beer-lambert}
 A=\varepsilon C L,
\end{equation}
yielding a value of
\begin{equation}
 \label{eq:QDconcentration}
 C_{QD}=(42 \pm 1)~\mathrm{nM}
\end{equation}
for the WbQD dispersion.

\subsection{Physical properties of CdS/ZnS QDs}
\label{sec:size}

The hydrodynamic size ($D_h$) of the CdS/ZnS QDs was measured using the dynamic light scattering (DLS) technique, in which intensity fluctuations caused by the scattering of laser radiation by freely diffusing QDs in a dispersion are auto-correlated~\cite{DLS-theory}. DLS measurements were performed on Zetasizer Nano ZS (Malvern), which performs measurements in triplicates and outputs the respective correlation curves, cumulant analysis results (mean particle size and standard deviation) and size distribution curves~\cite{DLS-cummulant-analysis}.

\begin{figure}[ht]
    \centering
\includegraphics{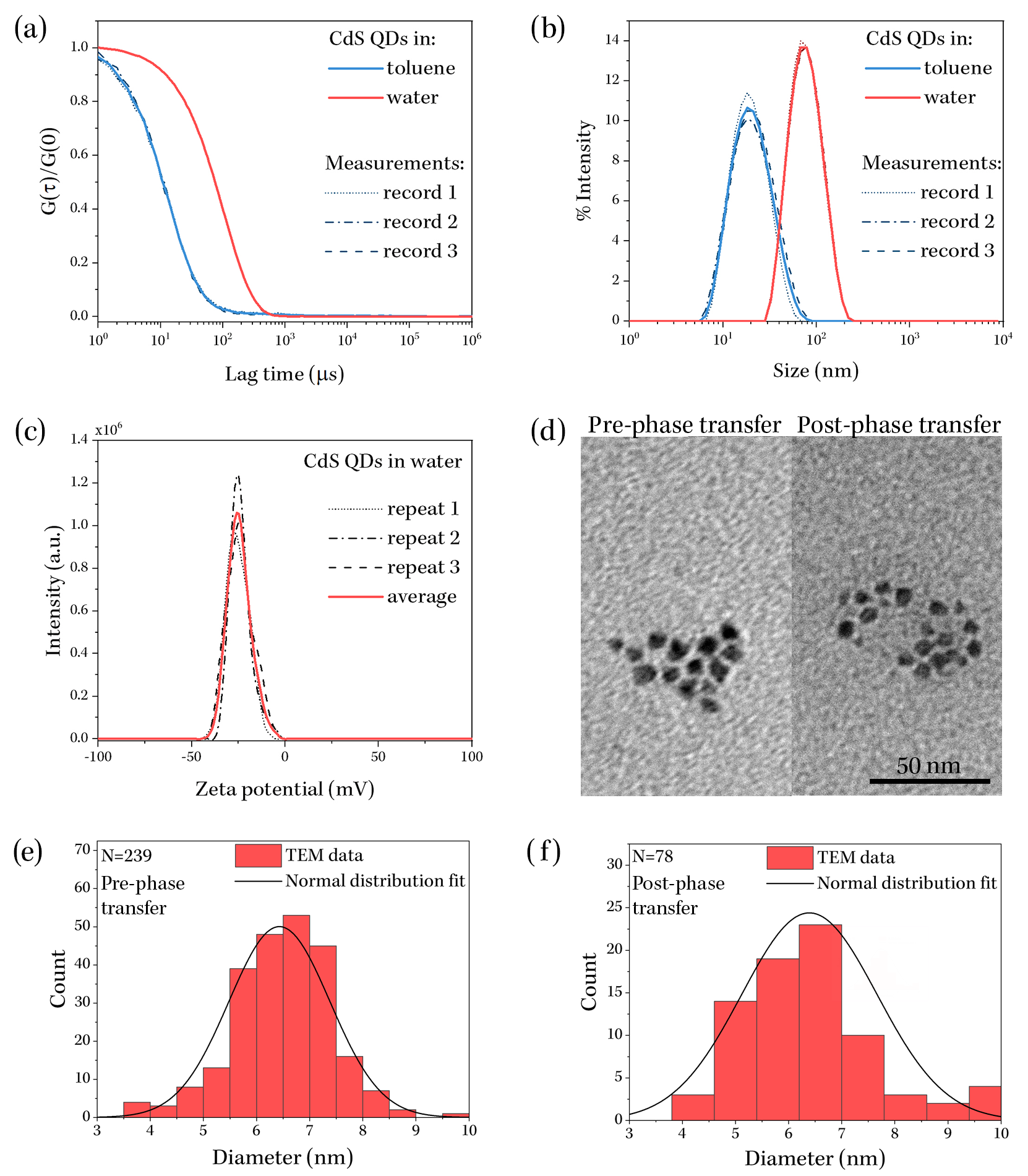}
\caption{\textbf{Measurements of QDs' size and zeta potential. (a)} Normalized correlation curves for CdS/ZnS QDs dispersed in toluene (blue) and in water (red). \textbf{(b)} Intensity hydrodynamic size distribution of CdS/ZnS QDs dispersed in toluene (blue) and in water (red), obtained from DLS measurements. \textbf{(c)} Intensity zeta potential distributions of CdS/ZnS QDs dispersed in water, measured using DLS measurements. In panels (a)-(c), the thinner non-solid lines show the repeats of individual measurements, whereas the thicker solid lines in brighter colour show the corresponding averages. \textbf{(d)} TEM images of the water-based CdS/ZnS QDs pre-phase transfer (left half) and post-phase transfer (right half). The scale bar is 50 nm long. Panels \textbf{(e)} and \textbf{(f)} show the distribution of QD diameters, obtained from TEM data (red bars), as well the normal distribution fit to the data (solid black line). Fitting was performed using OriginPro software, which also yielded mean and standard distribution of QDs sizes for each sample.} 
\label{fig:size-and-charge}
\end{figure}

Figure~\ref{fig:size-and-charge}(a) shows the normalized correlation curves of individual measurements as well as their average for CdS QDs dispersed in toluene and in water, after phase transfer. From this data, it is clear that correlation in scattered signals persists for longer lag times for WbQDs, which is indicative of their increased hydrodynamic size as compared to QDs dispersed in toluene. This was confirmed by the cumulant analysis, which yielded mean QD sizes of $\SizeDLSTC\pm\SizeDLSTE$~nm and $\SizeDLSWC\pm\SizeDLSWE$~nm pre- and post-phase transfer, with the population standard deviations of $\SizeDLSTS$~nm and $\SizeDLSWS$~nm respectively. The size distribution data, shown in Figure~\ref{fig:size-and-charge}(b) further confirms these results - the hydrodynamic size of WbQDs increases strongly compared to their predecessors dispersed in toluene. It is important to note that the observed large increase in the size of the QDs (approx. 3 fold) is due to both their increased physical size, owing to the additional OA layer around each QD, but also due to the emergence of charges on their surfaces which interact with the surrounding polar water molecules. It is known that the electrostatic force between charged particles and the polar water molecules contributes to the friction experienced by the particle when diffusing, resulting in larger effective hydrodynamic sizes~\cite{10.1063/1.434664,doi:10.1146/annurev.pc.31.100180.002021}. We confirmed the existence of surface charges on the phase-transfered QDs by measuring their zeta potential using DLS. Figure \ref{fig:size-and-charge}(c) shows the results of these measurements for WbQD sample. Analysis of these results (performed automatically by the Zetasizer software) yielded a mean zeta-potential value of
\begin{equation}
\zeta_{QD}=\Zetathrthr\pm\ZetathrthrE~\mathrm{mV} 
\end{equation}
for the sample. The polarity of this potential is consistent with that expected from QD functionalisation with the OA molecules. Furthermore, particles with zeta potentials of more than $+20$ mV or less than $-20$ mV tend to repel each other~\cite{https://doi.org/10.1111/jace.12371}; therefore, the value measured here is sufficient to ensure efficient dispersion of QDs \textit{via} electrostatic interactions between them and hence to ensure the colloidal stability of WbQDs.

Finally, the physical size of the QDs' inorganic layers (core + shell) was determined using Transmission Electron Microscopy (TEM), for both pre- and post-phaser transfer samples. Representative TEM images obtained from the measurements are shown in Fig.~\ref{fig:size-and-charge}(c). Analysis of TEM images yielded size distribution plots for each sample, shown in Figs.~\ref{fig:size-and-charge}(e) and (f). These data were fitted to a normal distribution model (usin OriginPro in-built fitting function) to obtain the mean QD diameter of pre-phase transfer QDs to be $\SizeTEMPT\pm\SizeTEMPTE$ nm and $\SizeTEMTC\pm\SizeTEMTE$ nm post-phase transfer. The two values were found to be equal, which is expected for the phase-transfer protocol employed ~\cite{oleic}, which has been previously shown to cause no change in the physical size of the QD inorganic core (including any inorganic shells) but a change in the overall size of the QDs due to the increased size of the OA layer surrounding each QD. Furthermore, both values are slightly larger but are comparable to the value obtained from the empirical model above ($\SizeEmpiC\pm\SizeEmpiE$ nm, using equation~\ref{eq:SizeEmpri}). This, again, is expected as the size of the QDs produced using the empirical model accounts only for the optically-active CdS core of the QDs and not the ZnS inorganic shell.

\subsection{Effects of sample ageing} \label{sec:aging}

Given that a typical neutrino experiment requires several years to acquire data, the degradation in scintillating properties of the fluors, caused by ageing, is a concern. For this reason, we performed some preliminary studies of the properties of WbQD samples over the span of 2 years. Three WbQD samples were prepared by the phase-transfer method for these purposes, in each of the years of 2021, 2022 and 2023, and their spectral and physical properties were measured.

\begin{figure}[ht]
    \centering
    \includegraphics{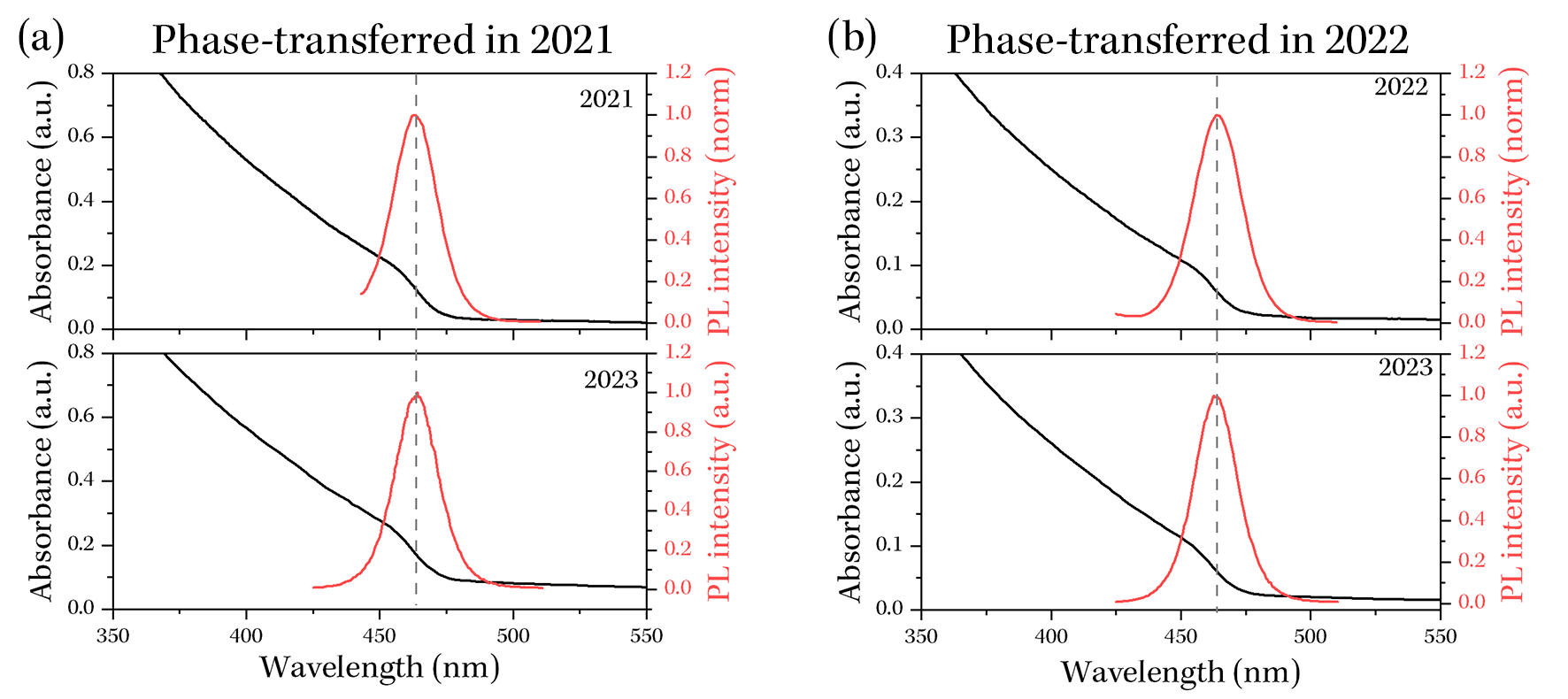}
    \caption{\textbf{Effects of sample ageing on the spectral properties of phase-transferred CdSe QDs.} Panels (a) and (b) show absorbance (black lines, left scales) and fluorescence (red lines, right scales) spectra of CdS QDs phase-transferred in July of \textbf{(a)} 2021 and \textbf{(b)} 2022. The top graphs in each panel show the optical spectra recorded for CdS samples immediately after phase-transfer process, and the bottom graphs show the same set of data measured in July of 2023. The gray dashed line in each panel indicates the peak wavelength of the freshly-transferred QDs' emission.}
    \label{fig:spectral-sample-aging}
\end{figure}

Figure~\ref{fig:spectral-sample-aging} compares the optical properties of the 2021 and 2022 WbQD samples, measured immediately after phase transfer in respective years and then in 2023, corresponding to 2 and 1 years of aging of the samples respectively. As can be seen from this figure, the photoluminescence and absorbance properties of these samples varied very little over time, with no observable changes or shifts in their spectral features. In all cases, the photoluminescence peaked in the $463-464$ nm range and any variations in the position of the peak could easily be explained by the changes in the environmental and experimental conditions under which measurements were performed, such as room temperature and humidity, as well as the calibration of the spectrometer's detector. That being said, for both samples, a strong change in the PLQY was observed from the moment of their initial preparation and measurements in 2023, with the PLQY of the 2021 sample decreasing from $\PLQYoneone\pm\PLQYoneoneE\%$ to $\PLQYonethr\pm\PLQYonethrE\%$ in the span of two years, whereas the PLQY of the 2022 sample increased from $\PLQYtwotwo\pm\PLQYtwotwoE\%$ to $\PLQYtwothr\pm\PLQYtwothrE\%$ over a 12 month period.

\begin{table}[ht]
\caption{Optical and physical properties of WbQD samples over time, showing the evolution of PLQY, hydrodynamic size ($D_h$) and the zeta potential ($\zeta$) of the WbQD samples right after phase transfer ("as prepared") and in 2023.}
\centering
\small
\begin{tabular}{>{\raggedright}p{0.12\textwidth}>{\centering}p{0.12\textwidth}>{\centering}p{0.2\textwidth}>{\centering}p{0.2\textwidth}>{\centering\arraybackslash}p{0.2\textwidth}}
 \hline
\textbf{Sample} & & PLQY (\%) & $D_h$ (nm) & $\zeta$ (mV) \\
 \hline
\multirow[m]{2}{*}{2021 WbQD} & as prepared & $\PLQYoneone\pm\PLQYoneoneE$ & $\Dhoneone\pm\DhoneoneE$ & not available \\
                              & in 2023 & $\PLQYonethr\pm\PLQYonethrE$ & $\Dhonethr\pm\DhonethrE$ & $\Zetaonethr\pm\ZetaonethrE$ \\\hline
\multirow[m]{2}{*}{2022 WbQD} & as prepared & $\PLQYtwotwo\pm\PLQYtwotwoE$ & $\Dhtwotwo\pm\DhtwotwoE$ & $\Zetatwotwo\pm\ZetatwotwoE$ \\
                              & in 2023 & $\PLQYtwothr\pm\PLQYtwothrE$ & $\Dhtwothr\pm\DhtwothrE$ & $\Zetatwothr\pm\ZetatwothrE$ \\\hline
2023 WbQD & as prepared & $\PLQYthrthr\pm\PLQYthrthrE$ & $\Dhthrthr\pm\DhthrthrE$ & $\Zetathrthr\pm\ZetathrthrE$\\
\hline
\end{tabular}
\label{tab:aging}
\end{table}

In view of the lack of any changes to the optical spectra of the samples, it is reasonable to conclude that these result imply that no long-term changes occur to the core of the QDs, since any such changes would cause shifts of spectral features; however, changes to the surface of the QDs - and namely the OA stabilizing shell - do occur, and they lead to changes in the PLQY of the QDs. Evidence to support the latter conclusion can be gained from the measurements of the WbQD samples' hydrodynamic sizes and zeta potentials ($\zeta$) over time. Table~\ref{tab:aging} shows the variations in these parameters over a period of 2 years. From the data presented in the table, no clear trend can be identified in the changes to the PLQY of the WbQDs samples: although the 2021 sample shows a reduction of PLQY after 2 years, the 2022 sample shows an opposite effect. On the other hand, $D_h$ consistently decreases over time for both samples, with no indication of agglomeration of colloidal QDs. Since the reduction is only a few nm, this is most likely due to the reorganization of OA the surface of WbQDs rather than their detachment/loss. Such surface reorganization would result in a change of $\zeta$ potential of the WbQDs, which was observed for the 2022 sample. However, since such data was not available for the as-prepared 2021 sample, the trend remains to be confirmed. In summary, whilst it is not clear what causes the changes in the properties of the WbQD samples from Table~\ref{tab:aging}, three things are confirmed: the hydrodynamic radius of the particles reduces gradually over time, WbQDs remain colloidally and optically stable over time, and no observable aggregation/precipitation of the sample over periods of time as large as two years post-transfer.

\section{Quantum Dots water solution cosmic ray test
\label{sec:cosmic}
}

\subsection{Setup}

\begin{figure}[ht]
    \centering
    \includegraphics{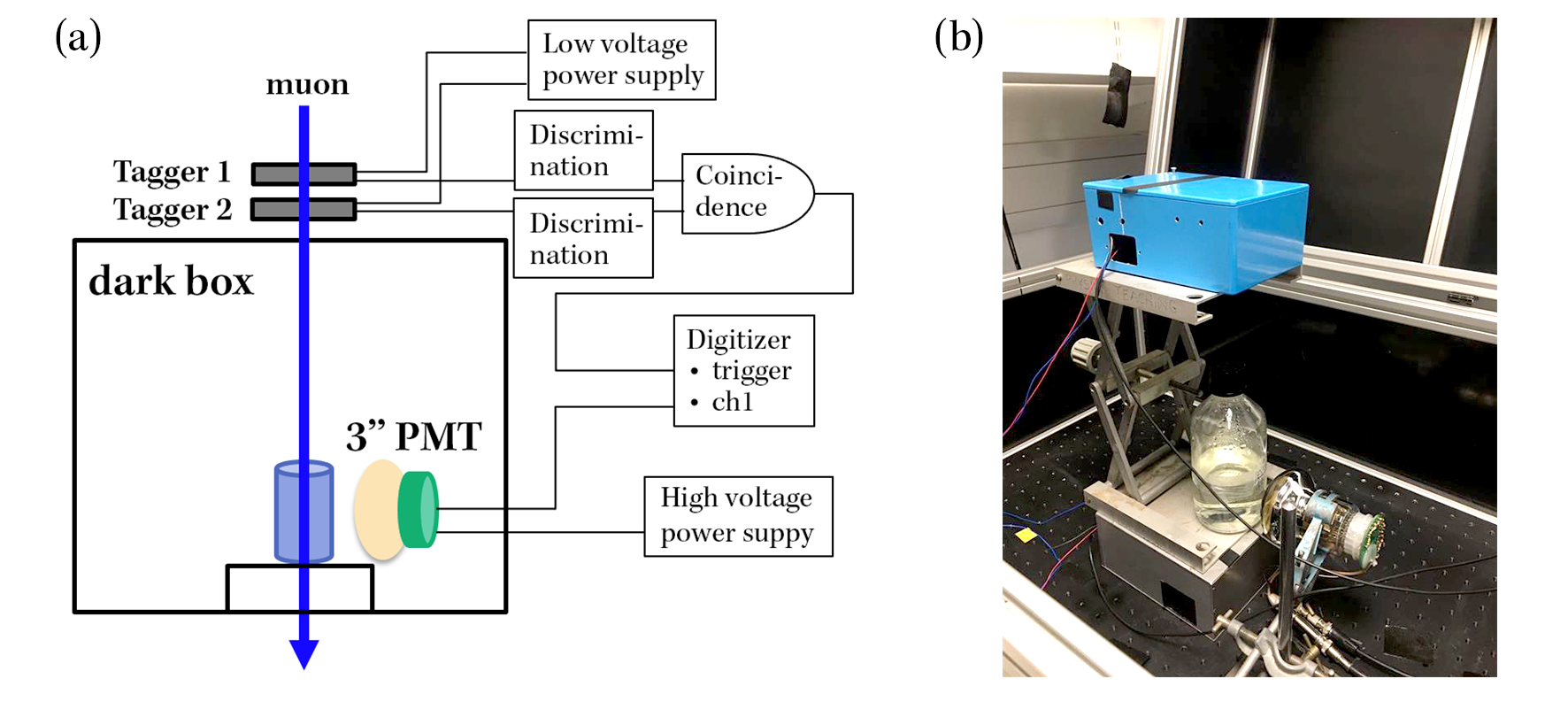}
    \caption{\textbf{(a)} Cosmic ray test diagram. 2 cosmic ray taggers are piled to define a cosmic ray trajectory. A bottle of CdS QDs in water solution is placed in a dark box where 3-inch PMT is located in front of it. Coincidence signals from the cosmic ray taggers are used to trigger the digitizer to record PMT pulses. \textbf{(b)} A view of the inside of the dark box. Note, cosmic ray tagger (the blue box) locates outside of the dark box unlike this picture.}
    \label{fig:cosmicraytest}
\end{figure}

The cosmic ray (atmospheric muon) measurement was performed in 2022 using the 2022 water-based CdS QDs sample. Cosmic rays were tagged by two cosmic ray taggers made with Hamamatsu 6~mm$^2$ silicon photo-multiplier (SiPM) C13367-6050EA~\cite{SiPM} and 100~cm$^2 \times 1$~cm EJ-200 plastic scintillator~\cite{Eljen}.
Fig.~\ref{fig:cosmicraytest} shows the setup. The QDs sample was placed in front of the
North Night Vision Technology (NNVT) negative high-voltage 3-inch photo-multiplier tube (PMT)~\cite{NNVT} in a dark box.
The signal and high voltage cables from the PMT were fed from the dark box to be connected with electronics outside. The signal cable was connected to the CAEN DT5720 250 MS/s digitizer~\cite{CAEN}. The signals from the cosmic ray taggers were processed using two discriminators connected to a coincidence unit, which produced logic signals to trigger the digitizer. 

\begin{figure}[ht]
    \centering
        \includegraphics{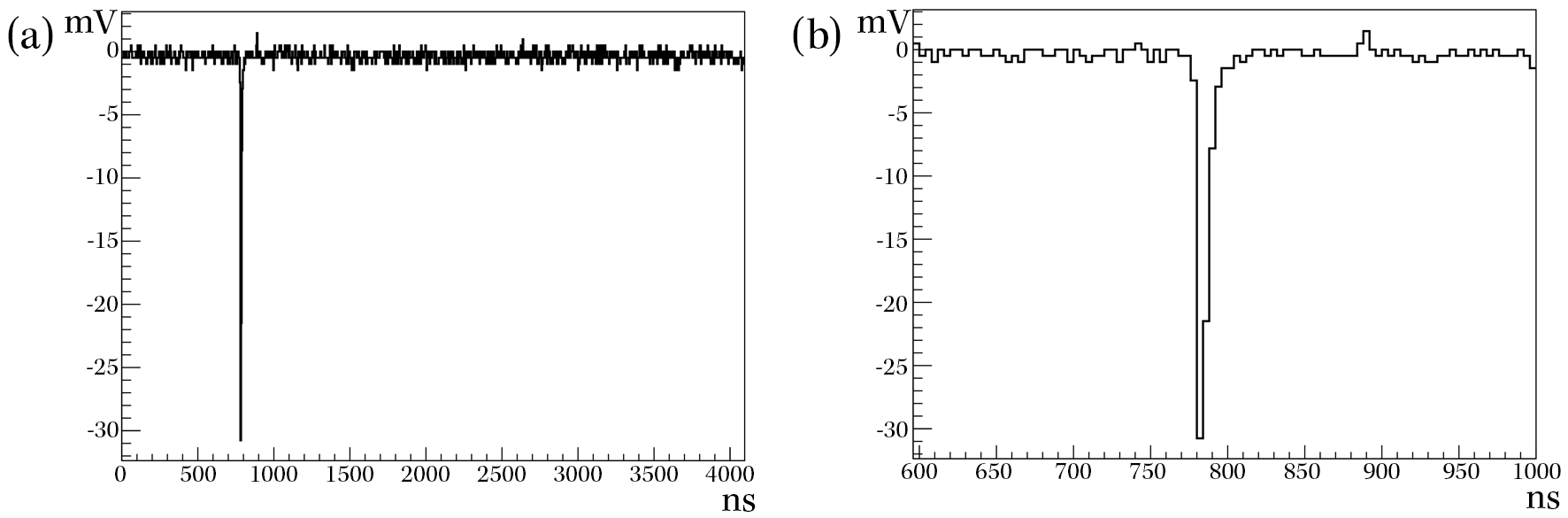}
    \caption{A typical PMT pulse recorded by the digitizer. Panel {\textbf{(a)}} shows the full data window and panel {\textbf{(b)}} shows the close-ups around the PMT pulse.}
    \label{fig:pulse}
\end{figure}

Fig.~\ref{fig:pulse} shows a typical recorded PMT pulse. Negative PMT pulses are recorded during the data acquisition window ($\sim\daqwindow~\mu$s).  In the offline analysis, the peak time is found by simple scanning, and charge integration is performed after the baseline subtraction. 

\begin{figure}[ht]
    \centering
\includegraphics{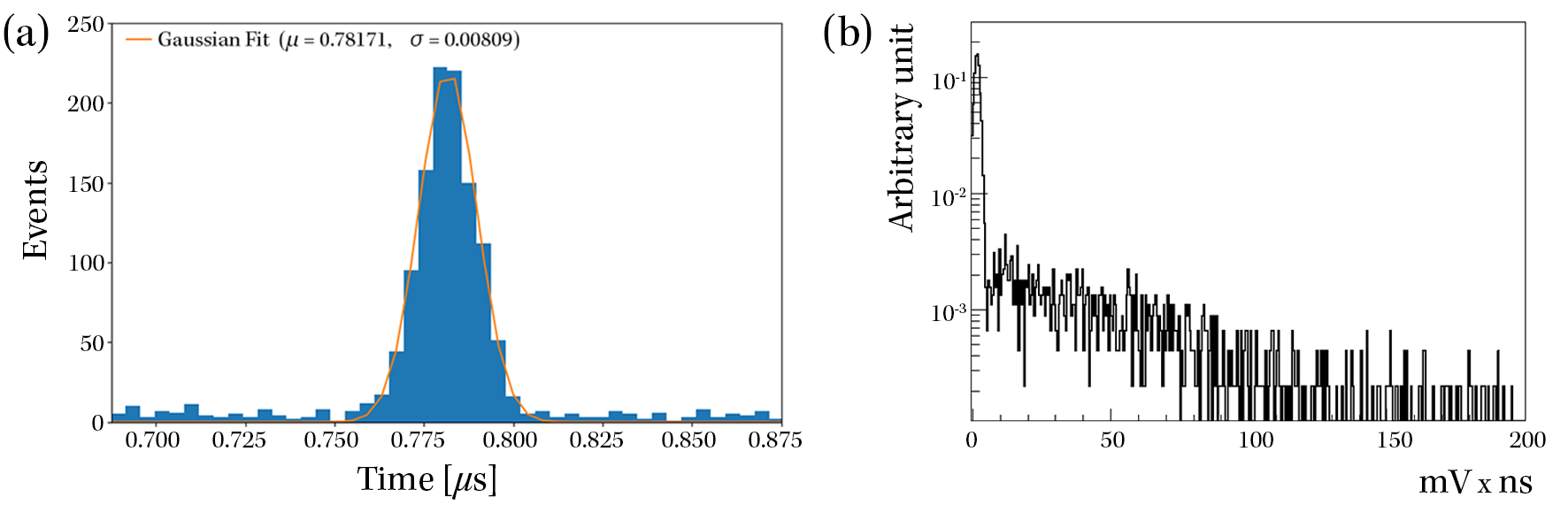}
    \caption{\textbf{(a)} QDs cosmic ray signal timing distribution.
\textbf{(b)} QDs cosmic ray signal charge distribution.}
    \label{fig:cosmicraydata}
\end{figure}

Fig.~\ref{fig:cosmicraydata} shows raw data distributions. The left plot shows the timing distribution of the pulse peak with an offset. Here, "t=0" is defined by the cosmic ray trigger signal. Since the QDs sample has only a few~cm depth in a bottle, it takes $<<1$~ns for cosmic rays to penetrate it. Thus the time distribution is limited by the decay time of QDs by excitation due to cosmic rays and the resolution. The Gaussian fit gives the decay time of QDs to be at most $\sim$8~ns, which is not atypical for a CdS QD sample ~\cite{QDlft1, QDlft2, QDlft3}. 
The right plot of Fig.~\ref{fig:cosmicraydata} shows the charge distribution in the unit of $mV\times ns$. The wide charge distribution indicates a wide range in both the angles at which the cosmic rays enter the QD sample, and in their propagation distances  once in the water media, resulting in deposition of a wide range of energies across the detected events. To better understand the measured data and their relation to the geometrical factors of the experimental set-up, a simulation was developed to estimate the numbers of photons of different energies that could be produced by the cosmic rays passing through our WbQD scintillating material.

\subsection{Simulation}

The simulation of this test was created in GEANT4~\cite{AGOSTINELLI2003250}, which is
a software toolkit for simulating physics processes relating
to particle propagation in matter utilising C++ object-oriented code.
The Geant4 toolkit contains applications and core physics header files,
which serves as the fundamental framework from which source files and other header files can be created.
These source files can be used to construct a vast array of experiments,
where the properties and the desired physical processes of the experimental apparatus can be tailored and produced, which in our case concerns a WbQD.
In this way, and with the inclusion of the detectors, the behaviour of the particles as they interact with the detector material can be simulated.

\begin{figure}[ht]
\centering
\includegraphics{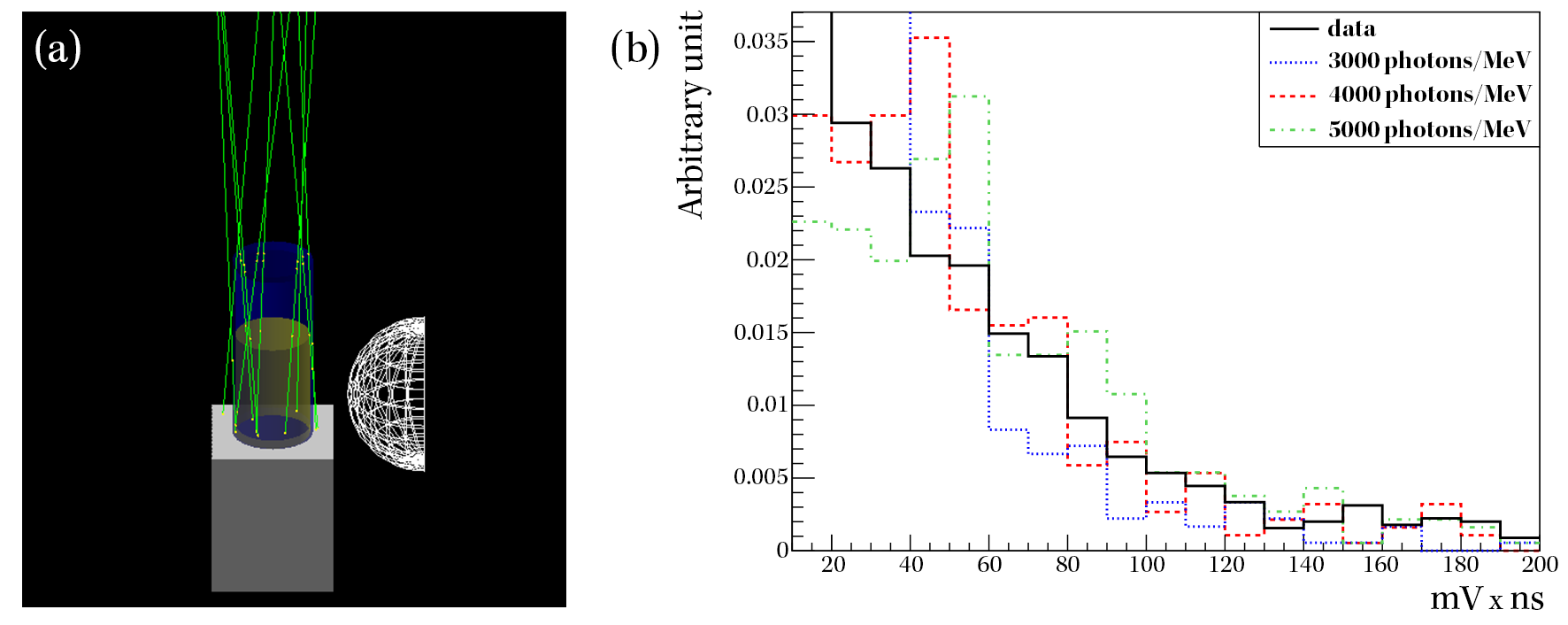}
\caption{\textbf{Simulation of the experimental setup in Geant4. (a)} Event display of the simulation. Cosmic ray distributions are based on the configuration of 2 cosmic ray taggers and example muon trajectories are shown in green. \textbf{(b)} Data-simulation charge distribution comparison. Data pedestal is cut, and simulation shows three predictions with the different number of scintillation photons per MeV, 3000~photons/MeV (blue dotted), 4000~photons/MeV (red dashed), and 5000~photons/MeV (green dash-dotted).}
\label{fig:sim}
\end{figure}

The geometry created is a simplified replica of the physical setup taking into consideration the distance, size, and shape of the objects.
Fig.~\ref{fig:sim} left shows an event display of our simulation in GEANT4.
The PMT hemispherical glass surface is set as the sensitive detector.
The WbQD material is treated as a liquid scintillator with close emission spectrum, located inside of a cylindrical glass.
Utilising the necessary GEANT4 libraries such as electromagnetic processes,
the physics processes can be constructed.
Muons are generated randomly and they penetrate the WbQDs volume
and produce scintillation photons that are detected by the PMT.
The trajectories of muons are calculated by the solid angle defined by 2 cosmic ray taggers, and for simplicity all muons are 4~GeV in this simulation.
In the simulation, we do not simulate PMT responses, instead, all photons which hit the surface of the PMT are registered with constant $15\%$ efficiency and the gain is $1.0\times 10^7$ at $-1100$~V according to our calibration.

Fig.~\ref{fig:sim} right shows data-MC comparison of their charge distributions above the pedestal. Here, the three simulations are presented with the different numbers of scintillation photons, and the agreement with the data is assessed from the tails of these distributions to match. A good agreement between experimental and simulated data is achieved around
$\QDSY$ photons/MeV, although with some deviations for low charge region. This is likely due to the contamination of pedestal events in the data, and further investigations are required to confirm this.
The simulation adopts a simple geometry, and objects in the dark box, except those shown in Fig.~\ref{fig:sim}, left, are ignored.
However, we believe they do not affect the time and charge distributions much, as major signals are direct light travelling from the WbQD scintillator to the front of the PMT. Importantly, this is the first demonstration of the water-based quantum dots liquid scintillator with radiation detection; furthermore, the number of photons recorded here is promising for applications in particle and nuclear physics.

\section{Discussions\label{sec:discussions}}

The results of the simulations indicate that several thousands of photons are created by CdS/ZnS QDs per 1 MeV of energy deposited, which is a surprisingly high number considering the low concentration of the dispersion ($42 \pm 1$ nM). For reference, a typical organic fluor in a liquid scintillator, such as 2,5-diphenyloxazole (PPO), would be present at a concentration of $4-30$ mM and would be expected to produce around 10,000 photons from 1 MeV energy deposit. The larger physical cross-sections of the QDs ($\approx32~\mathrm{nm^2}$ in our case), when compared to organic fluors ($0.5-1~\mathrm{nm^2}$), would be expected to somewhat mitigate for the lower concentration, but the differences in sizes are not sufficient to fully account for the observed photon rates. Furthermore, the PLQY of our QDs is also lower to that of a typical organic fluor ($9.5\%$ for WbQD sample and $85\%$ for PPO).

It is important to note that the nature of interactions that leads to the excitation of electron-hole pairs in the fluor is the same for both cases; however, the efficiency with which the initial excitation of electron-hole pairs can be achieved, the energy of the created pairs, and/or the processes that proceed this initial excitation are different for inorganic materials such as quantum dots. For example, the much larger number of atoms that constitute nanoparticulate fluors such as QDs, result in much higher electronic state densities being available for the transfer of energy when compared to organic fluors. As a result, excitations with a wide range of energies greater than the bandgap $E_g$ are possible in QDs, with those of energies $E>E_g$ leading to the generation of deep holes and hot electrons in the valence and conduction bands respectively. Furthermore, relaxation of these excitations, \textit{via} scattering, Auger and thermalization processes, generates either single or \emph{multiple} optically-gapped excitons at the band edge. Importantly, excitation of bi-excitons has been shown to be efficient in QDs and it can lead to the emission of two coincidental photons. In contrast, organic fluors do not offer this capability. High scintillation yields from QDs-based scintillators have been reported by other groups~\cite{Oktyabrsky:2016ard,10.1021/acsnano.0c06381,D1NA00815C}, suggesting that the optoelectronic properties of QDs offer advantages over organic compounds as scintillating materials. These results open up applications of liquid QD-based scintillators in particle physics~\cite{Winslow:2012ey,Blanco:2022cel,Doser:2022knm}.

\section{Conclusions\label{sec:conclusion}}

In this article, we investigate the potential of water-based quantum dots liquid scintillator for particle physics application. We first performed a phase transfer to CdS/ZnS QDs from hydrophobic phase to hydrophillic phase.
Absorbance and emission peaks for CdS QDs are $\sim\AbsWat$~nm and $\sim\EmiWat$~nm for both in toluene and in water. These values fit in the range provided by manufacturer, and we confirmed optical properties are mostly preserved in water-based quantum dots.

Photo luminescence quantum efficiency of the 2023 WbQD sample is measured to be $\PLQYthrthr\pm\PLQYthrthrE\%$. The value is lower than manufacture number, suggesting the loss due to phase transfer. Measured PLQY of other samples varies, and values go up or down over time and further study is needed to understand.

Hydrodynamic size of the toluene and water QD samples are measured to be $\SizeDLSTC\pm\SizeDLSTE$~nm and $\SizeDLSWC\pm\SizeDLSWE$~nm. TEM confirms the QD diameter to be $\SizeTEMTC\pm\SizeTEMTE$~nm, and a consistent diameter is obtained from an empirical equation using the absorption spectrum. Large WbQD size is attributed to the oleic acid layer with polar water molecules. This layer takes over 1 year to reorganize itself and it is stable over 2 years. Furthermore, there is no sign of agglomeration. The phase transfer has a significant impact for both PLQY and QD size, and it is a subject to be optimized in future.

Finally, we performed scintillation measurements from the cosmic rays (atmospheric muons). The time distribution confirms the fast decay ($<6$~ns) of QD is preserved in water, and charge distribution shows relatively high scintillation yield. More quantitative studies are needed to confirm these numbers.

In conclusion, water-based qunatum dots liquid scintillator is a promising new technology for particle physics. We have a special attention for a water solution of quantum dots to detect inverse beta decay processes from nuclear fission reactor neutrinos, where numerous programs are running all over the world~\cite{Bernstein:2019hix}. Although liquid scintillator detectors~\cite{Bernstein:2008tj,Classen:2015byu,NUCIFER:2015hdd,KamLAND:2010fvi,DoubleChooz:2019qbj,DayaBay:2022orm,JUNO:2021vlw,RENO:2020dxd,NEOS:2016wee,PROSPECT:2022wlf,STEREO:2022nzk,NEUTRINO-4:2018huq} are popular base design, other technologies such as solid scintillator~\cite{Oguri:2014gta,NuLat:2015wgu,Haghighat:2018mve,SoLid:2020cen} are also explored on top of water based detectors~\cite{Yeh:2011zz,Alonso:2014fwf,Theia:2019non,Caravaca:2020lfs,Kneale:2022vpw,Goldsack:2022vzm,SNO:2022qvw}. Furthermore, reactor neutrino detection from neutrino-electron elastic scattering~\cite{TEXONO:2009knm,Beda:2009kx} and neutrino-nucleus elastic scattering are established~\cite{Liao:2022hno,COHERENT:2023ffx}.

\section*{Acknowledgement}
We thank Mark Green for provision of characterisation equipment and of his expertise in phase-transfer protocols. We thank Stephen Po, Patrick Stowell, Klaus Suhling, and Lindley Winslow for provision of information and help. This paper is based on research funded by the ‘NuSec’ Nuclear Security Science Network managed by the University of Surrey and funded by the UKRI STFC (ST/S005684/1). Work by JC, SF, DM, AT are supported by the King’s Edge funding via King's Undergraduate Research Fellowships programme. Work by MT, NB, TK, are supported by the UKRI STFC (ST/W000660/1) and the Royal Society (RGS{\textbackslash}R2{\textbackslash}180163, IEC{\textbackslash}R3{\textbackslash}213009). Work by MZ, AR are supported by the UKRI EPSRC (EP/W017075/1) and the Royal Society (URF{\textbackslash}R{\textbackslash}211023, RF{\textbackslash}ERE{\textbackslash}231032).
\bibliographystyle{JHEP}
\bibliography{main}

\end{document}